\documentclass[apj,letterpaper]{emulateapj}
\usepackage{apjfonts}

\defcitealias{depropris99}{DP99}
\defcitealias{stanford02}{S02}
\defcitealias{bruzual03}{BC03}
\defcitealias{lin03b}{L03}
\defcitealias{lin04}{L04}
\slugcomment{ApJL, accepted}

\newcommand{\rmsub}[2]{\ensuremath{#1_{\mathrm{#2}}}} 
\newcommand{\eg}{{\textrm e.g.~}}

\newcommand\xray{\hbox{X--ray} }
\newcommand\beq{\begin{equation}}
\newcommand\eeq{\end{equation}}

\newcommand{\Msol}{\ensuremath{\mathrm{M}_{\odot}}}
\newcommand{\omegal}{\rmsub{\Omega}{\Lambda}}
\newcommand{\omegam}{\rmsub{\Omega}{M}}

\newcommand{\chandra}{\textit{Chandra}}

\newcommand{\xmme}{\textit{XMM}}

\shorttitle{Evolution of Cluster LF and Scaling Relations}
\shortauthors{Lin et al.}

\begin{document}

\title{Evolution of the $K$-band Galaxy Cluster Luminosity Function and
  Scaling Relations}

\author{
Yen-Ting Lin\altaffilmark{1,2,3},
Joseph J.~Mohr\altaffilmark{1,2,4},
Anthony H.~Gonzalez\altaffilmark{1,5},
and S.~Adam Stanford\altaffilmark{1,6}
}

\altaffiltext{1}{Visiting Astronomer, KPNO, NOAO.}
\altaffiltext{2}{Department of Astronomy, University of Illinois, Urbana, IL 61801}
\altaffiltext{3}{Current Address: 
 Departamento de Astronom\'{i}a y Astrof\'{i}sica, Pontificia Universidad Cat\'{o}lica de Chile;
 Princeton University Observatory, Princeton University, Princeton, NJ 08544; 
 ytlin@astro.princeton.edu}
\altaffiltext{4}{Department of Physics, University of Illinois,
 Urbana, IL 61801}
\altaffiltext{5}{Department of Astronomy, University of Florida,
  Gainesville, FL 32611}
\altaffiltext{6}{Department of Physics, University of California, Davis, CA 95616;
  IGPP, LLNL, Livermore, CA 94551}


\begin{abstract}

We study the evolution of two fundamental properties of galaxy clusters: the
luminosity function (LF) and the scaling relations between the total galaxy
number $N$ (or luminosity) and cluster mass $M$.  
Using a sample of 27 clusters
($0\le z \le 0.9$) with new near-IR observations and mass estimates derived from \xray 
temperatures, in conjunction with data from
the literature, we construct the largest sample for such studies to date.  The
evolution of the characteristic luminosity of the LF can be described by a
passively evolving population formed in a single burst at $z=1.5-2$. 
Under the assumption that the mass-temperature relation evolves self-similarly, and after the passive 
evolution is accounted for, the $N$--$M$ scaling shows no signs of evolution out to $z=0.9$.
Our data provide direct constraints
on halo occupation distribution models, and suggest that the way galaxies populate 
cluster-scale dark matter halos has not changed in the past 7 Gyr, in line with previous 
investigations.

\end{abstract}

\keywords{cosmology: observation -- galaxies: clusters: general 
  -- galaxies: luminosity function, mass function 
  -- galaxies: formation -- infrared: galaxies}

\section{Introduction}
\label{sec:intro}

The subject of cluster galaxy evolution has always been central to cluster
studies.  For the early types that comprise the majority of the cluster
galaxies, studies based on the fundamental plane and/or color-magnitude
relations suggest that their stars are primarily old, and evolve passively with
time \citep[\eg][]{bower92,stanford98,kelson00}.  Given the
hierarchical nature of structure formation, changes in the galaxy
population over cosmic time are expected.  Numerous investigations have
reported such evidence, \eg a higher fraction of blue galaxies in more distant
clusters \citep[\eg][]{butcher78,ellingson01}, a decline in the S0 galaxy
abundances towards higher-$z$ \citep{dressler97},
and a population that appears to be in the post-starburst phase
\citep{poggianti99}.

A powerful method for quantifying galaxy populations is via the luminosity
function (LF). The characteristic luminosity of the cluster galaxies as a whole
is reflected by the ``knee'' of the LF ($M_*$); comparison of LFs at different
redshifts thus will indicate the evolution of the mean galaxy population and
constrain the formation epoch of cluster galaxies
\citep[e.g.][]{barger98,depropris99,strazzullo06}.

Here we present an analysis of the evolution of the cluster $K_s$-band LF from $z=0$
to $z=0.9$, which has
several novelties and advantages over previous studies.  
Foremost, all our clusters have mass estimates accurate to $\lesssim 30\%$
(from the \xray temperature $T_X$).
We construct the LF
within the same portion of the cluster virial radius, which facilitates
comparisons of clusters at different redshifts and of different masses.  We
also make use of the cluster mass information and construct the LF in terms
of space density. 
Finally, our data generally are deeper, and cover a larger fraction of the
cluster virial region.  These aspects allow us to integrate the LF to
estimate the total galaxy luminosity ($L$) and number ($N$), and study their correlation
with the cluster mass ($M$).  Based on a large sample of nearby
clusters, we have found tight correlations between $L$ or $N$ with $M$
(\citealt{lin03b,lin04}, hereafter \citetalias{lin03b}, \citetalias{lin04}, respectively), 
which already provide an interesting constraint on the regularity 
of the cluster galaxy formation process.

In this study we pay special attention to the evolution of the $N$--$M$ relation, which
directly constrains simple clustering models such as the
halo occupation distribution (HOD; e.g.~\citealt{berlind02}) out to $z\sim 1$.
In such models, the way galaxies populate dark matter halos (e.g.~the number of 
cluster galaxies above some luminosity limit  
as a function of halo mass, known as the halo occupation number)
are tuned until the observed galaxy clustering at both small and large scales is 
reproduced.  Physics of galaxy formation is studied by identifying mechanisms that
reproduce the halo occupation statistics (e.g.~\citealt{zheng05}).
%
Furthermore, the evolution of the $N$--$M$ relation may reveal the mass function
evolution of subhalos in cluster-size halos \citep{cooray05a}.

In \S\ref{sec:data} we describe the sample selection, 
observations, data reduction, and our methods for the star-galaxy separation.
We stack the cluster data to construct the composite LFs at different
redshifts (\S\ref{sec:app}), which enable us to constrain the evolution of
$M_*$ and the formation epoch of the cluster galaxies.  We then proceed
to solve for the LF parameters of individual clusters, and examine the
evolution of the $L$--$M$ and $N$--$M$ relations (\S\ref{sec:sc}).  We discuss our
main findings in \S\ref{sec:sum}.  We assume the
cosmological parameters to be (\omegam, \omegal, $H_0$) = (0.3, 0.7,
$70\,h_{70}$\,km\,s$^{-1}$\,Mpc$^{-1}$). 
Throughout the paper we denote $K_s$-band by $K$-band for brevity, and
adopt the Vega magnitudes 
unless noted 
\citep[$K_{{\rm AB}}-K_{{\rm Vega}}=1.85$,][]{blanton06}.

\section{The Data}
\label{sec:data}

The cluster sample used here is drawn from several catalogs, and the selection
is heterogeneous. All our clusters have measured $T_X$, and have been observed
by either \chandra\ or \xmme.
The cluster near-IR (NIR) data were acquired with {\it FLAMINGOS} 
(hereafter FLMN) over several observing runs at 
the KPNO 4m and 2.1m telescopes, generally under $1.1\arcsec - 1.5\arcsec$
seeing.
Some of the basic information for the 27 clusters in our sample is listed in
Table~\ref{tab:1}, including: the cluster name, the redshift and
$T_X$ (in keV) of the cluster, the limiting $K$-band
magnitude, the fraction of the cluster virial
radius $r_{200}$\footnote{The virial mass is inferred from $T_X$ using
the empirical mass-temperature relation obtained by \citet{finoguenov01}. 
Self-similar evolution of the relation is assumed.
We denote the radius within which the mean
overdensity is $\Delta$ times the critical density as $r_\Delta$. 
The \citet[][NFW]{navarro97} profile with concentration $c=5$ is used for
converting between radii of different overdensities (\eg $r_{700}=0.56 r_{200}$,
$r_{1000}= 0.47 r_{200}$, $r_{2000}=0.33r_{200}$).
We use a $c=3$ profile for the spatial distribution of galaxies within clusters
\citepalias{lin04}.
}
that the FLMN field-of-view (FOV) corresponds to, and the available data in
other bands ($J$ and optical bands from the SDSS).  
The reduction of the data
was carried out by a pipeline developed at the University of Florida (see
\citealt{elston06}).

\begin{deluxetable}{lllccl}
\tablecolumns{6} 
\tablewidth{0pc}
\tabletypesize{\scriptsize}
\tablecaption{The Cluster Data\label{tab:1}}
\tablehead{
  \colhead{Name} & \colhead{$z$} &
  \colhead{$T_X$} & \colhead{$K_{lim}$}  &
  \colhead{$\theta_{max}/\theta_{200}$} & \colhead{Remarks\tablenotemark{a}} 
}
\startdata

Abell2052 & 0.0355 & \phn3.4  & 19.3 & 0.35 & 2m,S\\
Abell0085 & 0.0551 & \phn6.1 & 18.8 & 0.39 & 2m,S\\
Abell1795 & 0.0625 & \phn5.5 & 19.0 & 0.47 &  2m,S\\
Abell2029 & 0.0773 & \phn9.1 & 19.0 & 0.44 & 2m,S\\
Abell2255 & 0.0806 & \phn6.87 & 18.3 & 0.53 & 2m,S\\
Abell2142 & 0.0909 & \phn8.46 & 18.7 & 0.54  & 2m,S\\
Abell2244 & 0.0968 & \phn7.1 & 18.6 & 0.62 & 2m,S\\

Abell2055 & 0.1020 & \phn5.8 & 19.3 & 0.73 & 2m,S \\
Abell2034 & 0.1130 & \phn7.93 & 18.8 & 0.68 & 2m,S \\
Abell1413 & 0.1430 & \phn6.56 & 19.0 & 0.89 & 2m,S \\ 
Abell1204 & 0.1710 & \phn3.58 & 18.9 & 0.92 & 2m,S \\ 
Abell1914 & 0.1712 & \phn8.41 & 18.4 & 0.97 & 2m,S \\
Abell1689 & 0.1832 & \phn8.58 & 19.3& 1.02 & 2m,S\\

Abell2390  & 0.228 & 11.5  & 19.3 & 0.50 & 4m,J \\
RXJ2129.7+0005 & 0.234 & \phn6.78  & 19.4  & 0.68 & 4m,J,S \\
Abell1835      & 0.2532& \phn9.5   & 19.3 & 0.61 & 4m,J,S \\
Abell1758      & 0.279 & \phn6.57  &  18.6 & 1.70 & 2m,J,S \\
Abell1995      & 0.319 & 10.37 &  19.6  & 0.72 & 4m,J,S \\
MS1358.4+6245  & 0.328 & \phn6.92  & 20.4  & 0.91 & 4m,J,S \\
MS1621.5+2640  & 0.426 & \phn7.6   & 20.0  & 1.09 & 4m,J,S \\
RXJ1347.5$-$1145 & 0.451 & 14.1  & 18.8  & 0.83 & 4m,J \\
RXJ1701.3+6414 & 0.453 & \phn5.8   & 18.7  & 1.33 & 4m,S \\
RXJ1524.6+0957 & 0.516 & \phn5.1   & 19.8  & 1.59 & 4m,S \\
MACS0717.5+3745 & 0.548 & 14.4  &  19.3  & 0.97 & 4m,J \\
MACS0647.7+7015 & 0.584 & 11.58 &  19.0  & 1.16 & 4m,J \\
RXJ1120.1+4318 & 0.6   & \phn5.3   & 20.0  & 1.79 & 4m,J,S \\ 
MACS0744.8+3927 & 0.686 & 10.04 &  19.3  & 1.44 & 4m,J,S \\
\enddata

\tablenotetext{a}{2m \& 4m indicate the KPNO telescope used;
  J means $J$-band data also available; S means the cluster is covered by SDSS DR5.}
\end{deluxetable}

To increase the coverage in both redshift and mass, we supplement our FLMN data
with those presented in \citet[][hereafter
\citetalias{stanford02}]{stanford02}. The 45 clusters in
\citetalias{stanford02} were not selected by any well-defined criteria, and the
data were obtained with several different telescopes and instruments.
These instruments have a much smaller FOV than FLMN.  Some of these
data have been used by \citet[][hereafter
\citetalias{depropris99}]{depropris99} for the study of the $K$-band LF
evolution.  We include 14 clusters from \citetalias{stanford02} that have
measured $T_X$ and both $J$ and $K$-band data. 
The combined sample includes 41 clusters from $z\sim 0$ to $z\approx 0.9$,
all of which are observed out to at least $r_{2000} \approx r_{200}/3$
(Table~\ref{tab:1}).

The varying point spread function across the FLMN array prevents the use of 
morphological information for star-galaxy separation from our images.
For the majority of clusters at $z\ge 0.2$, both $J$ and
$K$-band photometry are available. In this case $J-K>1$ can
be used to select galaxies from the source catalog \citepalias{depropris99}.
For the clusters that lack
$J$-band data, we use data from the SDSS for help in star-galaxy
separation.  
Objects identified by the SDSS as point sources with $i_{AB}\le 21.3$ (the
95\% completeness limit for point sources) are regarded as stars.
Based on the $i-K$ color distribution of the sources 
from the FLAMINGOS Extragalactic Survey \citep[][hereafter
FLAMEX]{elston06}, we find that for objects brighter than $K =
18.06$, the SDSS identifies 95\% of the stars. For objects fainter than
$i_{AB} = 21.3$, we filter out the stars with the criterion  
$(g-i)_{{\rm AB}} \ge 1.4 (i-K)_{{\rm AB}} + 0.99$, 
which is a cut that we determine will filter 90\% of the stars.
At $K > 18.06$ the galaxy surface
density outnumbers that of the stars, and we conclude that our star-galaxy
separation scheme will induce $<1\%$ contamination in the final galaxy
catalogs.

\section{$K$-band Luminosity Functions from $\lowercase{z}=0$ to $\lowercase{z}=0.9$}
\label{sec:app}

The method developed in \citetalias{lin03b} for obtaining the total
galaxy luminosity or number of a cluster is to use the observed, 
background-corrected galaxy number
and flux to solve for the LF parameters $M_*$ and $\phi_*$, assuming a fixed
faint-end power-law slope $\alpha$ of the LF. 
Before we study the $N$--$M$ and $L$--$M$ relations at intermediate redshifts, it is
important to examine the evolution of the LF, as this (1) allows us to understand the
behavior of the faint-end, and (2) provides a means to apply the evolution correction
to the observed galaxy flux.

\begin{figure}
\epsscale{1.23}
\plotone{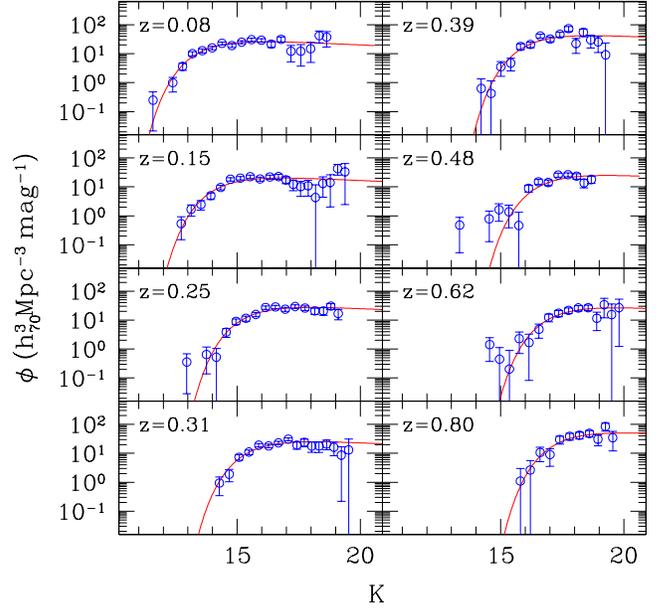}
\vspace{-8mm}
\caption{ 
  The evolution of apparent LFs, evaluated within $r_{2000}$.
  The LFs are presented in terms of the {\it physical} 
  space density (as opposed to the comoving density).
  The binning in redshift, and the number
  of clusters in each bin are: $0\le z <0.1$ (7),  $0.1\le z <0.2$ (6),  
  $0.2\le z <0.3$ (4),  $0.3\le z <0.34$ (8),  $0.34\le z <0.45$ (4),  
  $0.45\le z <0.55$ (3),  $0.55\le z <0.7$ (4),  $0.7\le z <0.9$ (5). 
  The midpoint redshifts are marked in each panel. 
  }
\label{fig:ap2000}
\end{figure}

Using the number counts derived from the FLAMEX survey for statistical background
correction, we follow \citetalias{lin04} to construct the composite LF in the observer's 
frame in eight redshift bins (Fig.~\ref{fig:ap2000}).
For each bin, we choose a midpoint redshift
$z_{cen}$, and adjust the $k$-correction and distance modulus of individual
clusters to best represent how the clusters would appear if they were all at 
$z_{cen}$ (e.g.~\citetalias{depropris99}). The combined corrections are typically 
less than $\pm 0.3$ mag.
As we discuss in \citetalias{lin04}, the main difference in the LFs between high and low 
mass clusters is the normalization $\phi_*$. Because our purpose here is to
examine the evolution of the LF shape, we ignore this difference in constructing the
composite LF.
In each panel we show the best-fit \citet{schechter76} function to the data, with $\alpha=
-0.9$. This value is consistent with what we find in \citetalias{lin04}, and
has been adopted by several previous studies.
We also note that the shape of the LFs ($K_*$ and the faint-end) does not change 
significantly within $30-60\%$ of 
$r_{200}$, which means the LFs shown in the Figure (evaluated within $r_{2000}$)
should be representative of the LF evolution.

\begin{figure}
\plotone{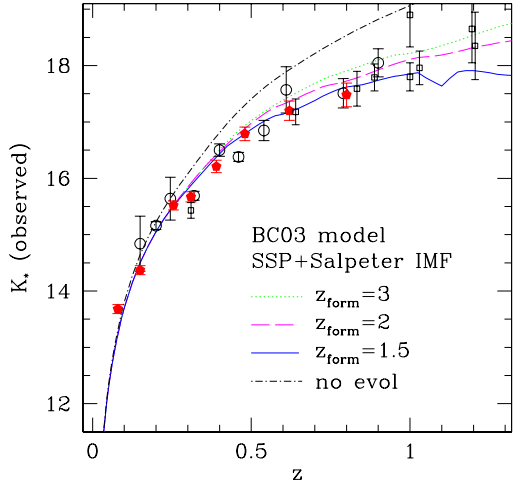}
\vspace{-6mm}
\caption{ 
  Evolution of $K_*$ as determined within $r_{2000}$ (solid pentagons).
  Results from \citetalias{depropris99} are shown as hollow circles. Other data points
  (open squares) are: two clusters at $z=0.31$ \citep{andreon05}, EIS0048 at
  $z=0.64$ \citep{massarotti03}, three clusters at $z=0.83$, 0.89, and 1.03
  \citep{ellis04}, two clusters at $z\sim 1$ \citep{kodama03}, MG2016+112 at
  $z=1$ \citep{toft03}, cluster 3C324 at $z=1.2$ \citep{nakata01}, 
 and three clusters at $z=1.2$ \citep{strazzullo06}.
}
\label{fig:ksev}
\end{figure}

In Fig.~\ref{fig:ksev} we show the best-fit $K_*$ as a function of redshift, and
compare them with the predictions from a population
synthesis model \citep[][hereafter \citetalias{bruzual03}]{bruzual03}. The
model galaxies are formed with the Salpeter initial mass function
and solar metallicity
in a single burst at $z_{{\rm form}} = 1.5$, 2, and 3. We have normalized
the model predictions to the $M_*$ found in 
\citetalias{lin04}.
The dot-dashed line in the Figure shows the $K_*$ with no evolution correction
applied.
We also show the results from several previous studies,
mainly that of \citetalias{depropris99}. 
Our data, together with most of the clusters at $z\le 1$ from the literature,
seem to favor the model where the stars in clusters are formed at $z_{{\rm
form}} = 1.5-2$.  None of the previous studies noted here examines the LF at a
fixed fraction of the virial radius. In the absence of any radial dependence of
the $M_*$ (see \eg \citetalias{lin04}), studying a region that is a constant
fraction of the virial region is not required.  Because we look at the cluster
galaxies as a whole, the behavior of $K_*(z)$ would also reflect the evolution
of field galaxies that fall into clusters much later than the old spheroids.
Our inferred $z_{{\rm form}}$ therefore seems low compared to that derived from
studies targeting early type cluster galaxies \citep[\eg][]{ellis97}.

\section{Evolution of the Near-IR Scaling Relations}
\label{sec:sc}

The results from the previous section indicate that (1) a passively evolving
stellar population formed in a single burst at $z=1.5$ can describe the LF (of
the whole cluster galaxy population) evolution based on our $z<0.9$ cluster
sample, and (2) the faint-end slope of LFs is consistent with $\alpha=-0.9$ at
all redshifts.  We can then use the \citetalias{bruzual03} model prediction of
the spectral aging for the $k$- and evolution corrections.  We follow
\citetalias{lin03b} to obtain the total galaxy luminosity and number for our
clusters.
Specifically, we integrate the LF over all galaxies more luminous than $M_{K*}(z)+2$,
where $M_{K*}(z)$ is the restframe characteristic magnitude based on the
\citetalias{bruzual03} model.

\begin{figure}
\epsscale{1.23}
\plotone{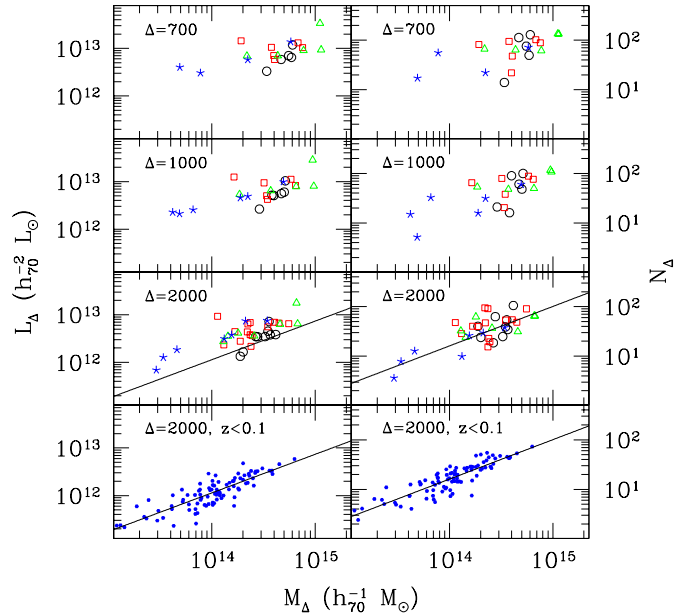}
\vspace{-8mm}
\caption{ 
  The panels on the left (right) column show the $L$--$M$ ($N$--$M$) correlation.
  The top three rows are the results measured 
  within $r_{700}$, $r_{1000}$, and $r_{2000}$, respectively. 
  Clusters in each redshift bin are plotted with different styles: 
  $z=0.0-0.2$: circle;
  $z=0.2-0.4$: square;       
  $z=0.4-0.6$: triangle;
  $z=0.6-0.9$: star.
  The bottom row shows the correlations for the $z<0.1$ {\it local} cluster sample
  (see text for details). The straight lines in the bottom two rows show the best-fit
  $L$--$M$ and $N$--$M$ correlations for the local sample.
}
\label{fig:scaling}
\end{figure}

We show in Fig.~\ref{fig:scaling} the $L$--$M$ and $N$--$M$ correlations evaluated
within $r_{700}$, $r_{1000}$, and $r_{2000}$. 
Clusters are binned into four redshift bins. 
It is interesting to see that these correlations also exist for the
intermediate redshift clusters studied here. 
For comparison, in the bottom panels are shown the correlations for the 93 clusters
at $z<0.1$ studied by \citetalias{lin04} (hereafter the {\it local} sample). 
For these local clusters $L$ and $N$ are both first 
determined within $r_{200}$ [down to $M_{K*}(z=0)+2$], then scaled to 
$r_{2000}$ using an NFW profile with $c=3$ (e.g.~\citetalias{lin04}).

\begin{figure}
\plotone{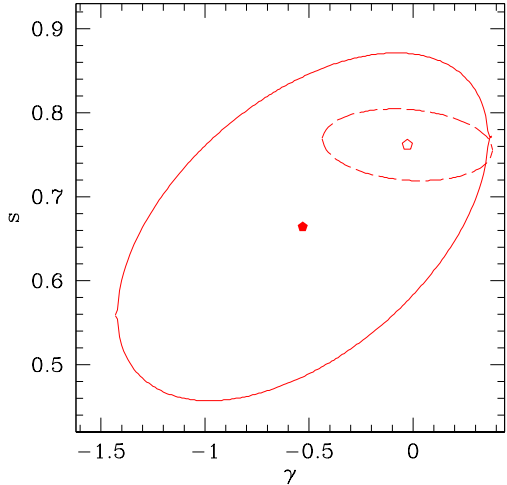}
\vspace{-6mm}
\caption{ 
  Joint 68\% confidence region for $\gamma$ and $s$
  (see Eqn.~\ref{eq:sc} for the definition). 
  The large contour results when Eqn.~\ref{eq:sc} is fit to the high-$z$ sample. 
  A tighter constraint on $\gamma$ is obtained 
  when both high-$z$ and local samples are used (the dashed 
  contour). The points show the
  best-fit values for ($\gamma$, $s$): ($-0.53$, 0.66) for the high-$z$ sample, and
  ($-0.03$, 0.76) for the combined sample.
}
\label{fig:evol}
\end{figure}

We examine the existence and the form of evolution in the data by
considering \beq \label{eq:sc} N(M,z) = N_0 (1+z)^\gamma
(M/M_0)^s \eeq for the $N$--$M$ evolution, where $N_0$ and $M_0$ are the
normalization factors of the relation. The evolution is specified by 
$\gamma$. With $M_0$ set at $10^{14.3} \Msol$, we fit the above
expression to all clusters to find out the parameters ($N_0$, $\gamma$, $s$).
It is expected that the parameters $\gamma$ and $s$ would exhibit some
degeneracy, and therefore we also include the local sample for this exercise,
whose size and the range in mass would
greatly help determine $N_0$ and $s$.
%
%
Within different regions (e.g.~$r_{700}$, $r_{1000}$, and $r_{2000}$), the constraints
on the evolution of the $N$--$M$ relation are consistent.
For simplicity, we only present the results from
$N_{2000}$--$M_{2000}$ correlations, which are the tightest of the three cases.
In Fig.~\ref{fig:evol} we show the joint constraints for $\gamma$ and $s$. 
The smaller (larger) contour is obtained when the local sample is included 
(excluded). The contours correspond to 68\% confidence region. 
For each case the best-fit
($\gamma$, $s$) is shown as the symbol at the center of the contour. 
Combining both samples results in a much tighter constraint on $\gamma$.
If we further marginalize
over $s$, we find that without the local sample, $\gamma = -0.53 \pm 0.59$;
inclusion of the local sample gives $\gamma=-0.03\pm 0.27$. Both results are
consistent with a no-evolution scenario.

\section{Discussion}
\label{sec:sum}

We can understand the no-evolution of the $N$--$M$ relation by looking at the 
integration of the LF:
\beq
N(M,z) = V \phi_* \Gamma(\alpha+1, L_{l}(z)/L_*(z)),
\eeq
where $V$ is the cluster volume, $L_{l}$ is the luminosity corresponding to the
lower limit of the integration, and $\Gamma$ is the incomplete gamma function.
Because $L_l$ is set to be a fixed fraction of $L_*$, it is the product of $V$
and $\phi_*$ that determines $N$. For two clusters $i$ and $j$ which have
the same mass but at different redshifts (say $z_j>z_i$), $V_j<V_i$ (assuming that
$r_\Delta \propto 1/H(z)$, where $H(z)$ is the expansion rate of the Universe).
One thus expects that on average, $\phi_{*j}>\phi_{*i}$. 
Indeed, $\phi_*$ derived for our clusters agrees with this expectation.
{\it That the increase in the galaxy number density is offset by the decrease in the virial
volume is the main reason for the lack of evolution of the $N$--$M$ relation.}

Previous studies have found suggestions of lack of HOD evolution toward $z\sim
1$, based on clustering statistics \citep{yan03,phleps06,coil06}.
In a recent
attempt to extract both cosmological information and cluster physics from a
large sample of clusters, it is found that the optical observable--mass
relation, although only weakly constrained, does not evolve with redshift
\citep{gladders06}. We note, however, passive evolution has not been 
accounted for in this analysis (S.~Majumdar 2006, private communication).  
Our study is complementary in the sense
that we probe the (high mass end of) occupation number directly, with
reliable cluster mass estimates from $T_X$.

It is possible that changes in both the red and blue galaxy populations with redshift
are hidden from the evolution of the $N$--$M$ relation, if they conspire to make the total
galaxy number unchanged. We are in the process of acquiring color 
information in order to constrain the evolution of these
populations separately. 

The fact that only when the large, nearby cluster sample of
\citetalias{lin04} is included can we obtain reasonably tight constraint on the
$N$--$M$ evolution points to the need for a {\it much} larger sample with reliable mass
estimates at
higher-$z$. This is also demanded for large optical/NIR cluster surveys which
may rely on optical/NIR based cluster mass estimators (\eg FLAMEX).  
Based on simple Monte Carlo simulations, we estimate that at least 20 clusters at $z=0.8-1.0$ 
are required to constrain $\gamma$ to 0.1 level.
While more clusters at intermediate redshifts may provide better constraints on the
evolution of the slope of the scaling relation, $z>1$ clusters clearly will help pinpoint the
formation epoch of the cluster galaxies.

\acknowledgments 

This paper is dedicated to the
memory of Richard Elston, whose instrument, {\it FLAMINGOS}, and observation
programs, are essential to this analysis.  
We acknowledge Z.~Zheng, A.~Sanderson, D.~Nagai, 
and D.~Stern for useful
comments on the manuscript.  
D.~Stern is also thanked for help in data acquisition.  
We thank an anonymous referee for helpful suggestions.
JJM and YTL acknowledge
support from NSF OPP award OPP-0130612 and NASA LTSA award
NAG5-11415.  
YTL is also supported by NSF PIRE grant OISE-0530095.  AHG
acknowledges support from NSF award AST-0436681.  SAS's work was performed under 
the auspices of the U.S.~DOE,
NNSA by the UC,
LLNL under contract No.~W-7405-Eng-48.
This work uses data acquired as part of the FLAMEX program.
This publication has made use of data products from the SDSS. Its website is http://www.sdss.org/.


\begin{thebibliography}{36}
\expandafter\ifx\csname natexlab\endcsname\relax\def\natexlab#1{#1}\fi

\bibitem[{{Andreon} {et~al.}(2005){Andreon}, {Punzi}, \& {Grado}}]{andreon05}
{Andreon}, S., {Punzi}, G., \& {Grado}, A. 2005, \mnras, 360, 727

\bibitem[{{Barger} {et~al.}(1998){Barger}, {Aragon-Salamanca}, {Smail},
  {Ellis}, {Couch}, {Dressler}, {Oemler}, {Poggianti}, \&
  {Sharples}}]{barger98}
{Barger}, A.~J., et~al. 1998, \apj, 501, 522


\bibitem[{{Berlind} \& {Weinberg}(2002)}]{berlind02}
{Berlind}, A.~A. \& {Weinberg}, D.~H. 2002, \apj, 575, 587

\bibitem[{{Blanton} \& {Roweis}(2006)}]{blanton06}
{Blanton}, M.~R. \& {Roweis}, S. 2006, \aj, submitted (astro-ph/0606170)

\bibitem[{{Bower} {et~al.}(1992){Bower}, {Lucey}, \& {Ellis}}]{bower92}
{Bower}, R.~G., {Lucey}, J.~R., \& {Ellis}, R.~S. 1992, \mnras, 254, 601

\bibitem[{{Bruzual} \& {Charlot}(2003)}]{bruzual03}
{Bruzual}, G. \& {Charlot}, S. 2003, \mnras, 344, 1000 (BC03)

\bibitem[{{Butcher} \& {Oemler}(1978)}]{butcher78}
{Butcher}, H. \& {Oemler}, A. 1978, \apj, 219, 18

\bibitem[{{Coil} {et~al.}(2006){Coil}, {Gerke}, {Newman}, {Ma}, {Yan},
  {Cooper}, {Davis}, {Faber}, {Guhathakurta}, \& {Koo}}]{coil06}
{Coil}, A.~L., et~al. 2006, \apj, 638, 668


\bibitem[{{Cooray} \& {Milosavljevi{\'c}}(2005)}]{cooray05a}
{Cooray}, A. \& {Milosavljevi{\'c}}, M. 2005, \apjl, 627, L89

\bibitem[{{De Propris} {et~al.}(1999){De Propris}, {Stanford}, {Eisenhardt},
  {Dickinson}, \& {Elston}}]{depropris99}
{De Propris}, R., {Stanford}, S.~A., {Eisenhardt}, P.~R., {Dickinson}, M., \&
  {Elston}, R. 1999, \aj, 118, 719 (DP99)

\bibitem[{{Dressler} {et~al.}(1997){Dressler}, {Oemler}, {Couch}, {Smail},
  {Ellis}, {Barger}, {Butcher}, {Poggianti}, \& {Sharples}}]{dressler97}
{Dressler}, A., et~al. 1997,
  \apj, 490, 577


\bibitem[{{Ellingson} {et~al.}(2001){Ellingson}, {Lin}, {Yee}, \&
  {Carlberg}}]{ellingson01}
{Ellingson}, E., {Lin}, H., {Yee}, H.~K.~C., \& {Carlberg}, R.~G. 2001, \apj,
  547, 609

\bibitem[{{Ellis} {et~al.}(1997){Ellis}, {Smail}, {Dressler}, {Couch},
  {Oemler}, {Butcher}, \& {Sharples}}]{ellis97}
{Ellis}, R.~S., {Smail}, I., {Dressler}, A., {Couch}, W.~J., {Oemler}, A.~J.,
  {Butcher}, H., \& {Sharples}, R.~M. 1997, \apj, 483, 582

\bibitem[{{Ellis} \& {Jones}(2004)}]{ellis04}
{Ellis}, S.~C. \& {Jones}, L.~R. 2004, \mnras, 348, 165

\bibitem[{{Elston} {et~al.}(2006){Elston}, {Gonzalez}, {McKenzie}, {Brodwin},
  {Brown}, {Cardona}, {Dey}, {Dickinson}, {Eisenhardt}, {Jannuzi}, {Lin},
  {Mohr}, {Raines}, {Stanford}, \& {Stern}}]{elston06}
{Elston}, R.~J., {Gonzalez}, A.~H., et~al. 2006, \apj, 639, 816


\bibitem[{{Finoguenov} {et~al.}(2001){Finoguenov}, {Reiprich}, \&
  {B{\"o}hringer}}]{finoguenov01}
{Finoguenov}, A., {Reiprich}, T.~H., \& {B{\"o}hringer}, H. 2001, \aap, 368,
  749

\bibitem[{{Gladders} {et~al.}(2006){Gladders}, {Yee}, {Majumdar}, {Barrientos},
  {Hoekstra}, {Hall}, \& {Infante}}]{gladders06}
{Gladders}, M.~D., {Yee}, H.~K.~C., {Majumdar}, S., {Barrientos}, L.~F.,
  {Hoekstra}, H., {Hall}, P.~B., \& {Infante}, L. 2006, \apj, submitted
  (astro-ph/0603588)


\bibitem[{{Kelson} {et~al.}(2000){Kelson}, {Illingworth}, {van Dokkum}, \&
  {Franx}}]{kelson00}
{Kelson}, D.~D., {Illingworth}, G.~D., {van Dokkum}, P.~G., \& {Franx}, M.
  2000, \apj, 531, 184

\bibitem[{{Kodama} \& {Bower}(2003)}]{kodama03}
{Kodama}, T. \& {Bower}, R. 2003, \mnras, 346, 1

\bibitem[{{Lin} {et~al.}(2003){Lin}, {Mohr}, \& {Stanford}}]{lin03b}
{Lin}, Y.-T., {Mohr}, J.~J., \& {Stanford}, S.~A. 2003, \apj, 591, 749 (L03)

\bibitem[{{Lin} {et~al.}(2004){Lin}, {Mohr}, \& {Stanford}}]{lin04}
---. 2004, \apj, 610, 745 (L04)

\bibitem[{{Massarotti} {et~al.}(2003){Massarotti}, {Busarello}, {La Barbera},
  \& {Merluzzi}}]{massarotti03}
{Massarotti}, M., {Busarello}, G., {La Barbera}, F., \& {Merluzzi}, P. 2003,
  \aap, 404, 75

\bibitem[{{Nakata} {et~al.}(2001){Nakata}, {Kajisawa}, {Yamada}, {Kodama},
  {Shimasaku}, {Tanaka}, {Doi}, {Furusawa}, {Hamabe}, {Iye}, {Kimura},
  {Komiyama}, {Miyazaki}, {Okamura}, {Ouchi}, {Sasaki}, {Sekiguchi}, {Yagi}, \&
  {Yasuda}}]{nakata01}
{Nakata}, F., et~al. 2001, \pasj, 53, 1139



\bibitem[{{Navarro} {et~al.}(1997){Navarro}, {Frenk}, \& {White}}]{navarro97}
{Navarro}, J.~F., {Frenk}, C.~S., \& {White}, S. D.~M. 1997, \apj, 490, 493

\bibitem[{{Phleps} {et~al.}(2006){Phleps}, {Peacock}, {Meisenheimer}, \&
  {Wolf}}]{phleps06}
{Phleps}, S., {Peacock}, J.~A., {Meisenheimer}, K., \& {Wolf}, C. 2006, \aap,
  accepted (astro-ph/0506320)

\bibitem[{{Poggianti} {et~al.}(1999){Poggianti}, {Smail}, {Dressler}, {Couch},
  {Barger}, {Butcher}, {Ellis}, \& {Oemler}}]{poggianti99}
{Poggianti}, B.~M., {Smail}, I., {Dressler}, A., {Couch}, W.~J., {Barger},
  A.~J., {Butcher}, H., {Ellis}, R.~S., \& {Oemler}, A.~J. 1999, \apj, 518, 576

\bibitem[{Schechter}(1976)]{schechter76} {Schechter}, P.~L. 1976, \apj, 203, 297


\bibitem[{{Stanford} {et~al.}(1998){Stanford}, {Eisenhardt}, \&
  {Dickinson}}]{stanford98}
{Stanford}, S.~A., {Eisenhardt}, P.~R., \& {Dickinson}, M. 1998, \apj, 492, 461

\bibitem[{{Stanford} {et~al.}(2002){Stanford}, {Eisenhardt}, {Dickinson},
  {Holden}, \& {De Propris}}]{stanford02}
{Stanford}, S.~A., {Eisenhardt}, P.~R., {Dickinson}, M., {Holden}, B.~P., \&
  {De Propris}, R. 2002, \apjs, 142, 153

\bibitem[{{Strazzullo} {et~al.}(2006){Strazzullo}, {Rosati}, {Stanford},
  {Lidman}, {Nonino}, {Demarco}, {Eisenhardt}, {Ettori}, {Mainieri}, \&
  {Toft}}]{strazzullo06}
{Strazzullo}, V., et~al. 2006, \aap, 450, 909


\bibitem[{{Toft} {et~al.}(2003){Toft}, {Soucail}, \& {Hjorth}}]{toft03}
{Toft}, S., {Soucail}, G., \& {Hjorth}, J. 2003, \mnras, 344, 337

\bibitem[{{Yan} {et~al.}(2003){Yan}, {Madgwick}, \& {White}}]{yan03}
{Yan}, R., {Madgwick}, D.~S., \& {White}, M. 2003, \apj, 598, 848

\bibitem[{{Zheng} {et~al.}(2005){Zheng}, {Berlind}, {Weinberg}, {Benson},
  {Baugh}, {Cole}, {Dav{\'e}}, {Frenk}, {Katz}, \& {Lacey}}]{zheng05}
{Zheng}, Z., et~al. 2005, \apj, 633, 791


\end{thebibliography}
\end{document}